\documentclass[12pt]{iopart}

\usepackage{iopams}
\usepackage{graphicx}
\usepackage{amsfonts}
\usepackage{amsbsy}
\usepackage{xcolor}
\usepackage{verbatim}

\pdfoutput=1
\renewcommand{\vec}[1]{\mbox{\boldmath$\mathrm{#1}$}}
\def\ind#1{{_{\mathrm{#1}}}}

\begin{document}

\title[Dipole-Dipole Interaction in arrays of Fe/Fe$_{\mathrm{x}}$O$_{\mathrm{y}}$ core/shell nanocubes probed by FMR]{Dipole-Dipole Interaction in arrays of Fe/Fe$_{\mathrm{x}}$O$_{\mathrm{y}}$ core/shell nanocubes probed by ferromagnetic resonance}

\author{A. Sukhov$^1$, P.P. Horley$^2$, J. Berakdar$^1$, A. Terwey$^3$, R. Meckenstock$^3$, M. Farle$^3$}

\address{$^1$Institut f\"ur Physik, Martin-Luther-Universit\"at Halle-Wittenberg, 06120 Halle (Saale), Germany\\
$^2$Centro de Investigaci\'{o}n en Materiales Avanzados (CIMAV S.C.), Chihuahua/Monterrey, 31109 Chihuahua, Mexico\\
$^3$Faculty of Physics and Center for Nanointegration (CeNIDE), University of Duisburg-Essen, 47057 Duisburg, Germany}


\begin{abstract}
This paper represents a detailed theoretical study of the role of the long-range magnetic dipole-dipole interaction evidenced by the ferromagnetic resonance (FMR) spectra for the ordered arrays of cubic nanoparticles. We show that the size of the array essentially controls the stability of the system, allowing to suppress the intermittent low-field excitations starting from the arrays formed by $6 \times 6$ nanoparticles. Our numerical simulations allow to determine the threshold inter-particle distance (around $80\div 100$ nm), after which the dipole-dipole interaction becomes negligible so that the FMR spectrum of the nanoparticle arrays becomes the same as the spectrum featured by a single nanoparticle. \textcolor{black}{ We also compare our simulations with experimental FMR-spectra of 24 Fe/Fe$_{\mathrm{x}}$O$_{\mathrm{y}}$-nanocubes irregularly placed on a substrate.}
\end{abstract}


\maketitle

\section{Introduction}

During several last decades new fabrication and characterization techniques \cite{ZaFa13} fostered the discovery of new properties of materials on the scale from several hundreds- to several nanometers. In view of miniaturization of all devices used in the information technology and in the information storage in particularly \cite{WeMo99} it is desired to design nanoscale magnetic objects\textcolor{black}{, where bits of information could be stored with a well defined temporally stable direction of the magnetization}. Spherical ferromagnetic Fe$_{\mathrm{x}}$Pt$_{\mathrm{1-x}}$-nanoparticles \cite{SuMu00,DoFi97,CoKo99,AnLi05} \textcolor{black}{ and also FePt$_3$-nanocubes \cite{MaTr07} were considered as likely objects due to their large magnetocrystalline anisotropy. The design of a unique direction of the magnetization in arrays of spherical particles, however, turned out to be impossible so far}. \textcolor{black}{ The formation of FePt/Fe$_{\mathrm{x}}$O$_{\mathrm{y}}$ nanocubes might provide a feasible route, which requires the understanding of the high-frequency switching behavior of arrays of such cubes and the role of dipolar interactions on the damping mechanisms.} For Fe/Fe$_{\mathrm{x}}$O$_{\mathrm{y}}$-nanocubes \cite{ShRo07} \textcolor{black}{ the advantage of their controlled placement on the substrate allows for a dramatic reduction of the parameter space in the interpretation of resonance spectra.}

Fe/Fe$_{\mathrm{x}}$O$_{\mathrm{y}}$-nanocubes can be chemically prepared \cite{Terw12,TrMe08} from iron stearate and sodium oleate via heating and a subsequent cooling down. These ferromagnetic (FM) nanoparticles can form two dimensional arrays on a silicon-oxide substrate. Further analysis shows that the nanoparticles consist of a bcc Fe-core and a Fe$_{\mathrm{x}}$O$_{\mathrm{y}}$-shell (around $7$ nm thick \cite{Terw12}), implying that 66$\%$ of all atoms are in the shell and only 34$\%$ are in the volume (\cite{Terw12} p. 22).

Transmission electron microscope (TEM) images \cite{Terw12} and \textcolor{black}{ 3D tomography reveal the nearly perfect cubic form of the nanoparticles}. Statistical analysis for such nanoparticles yielded the average lateral size of around $43$ nm \cite{Terw12}.

In a recent study \cite{KrFr11} of Fe/Fe$_{\mathrm{x}}$O$_{\mathrm{y}}$-nanocubes of different size, the influence of the nanoparticles' arrangement on the magnetic hysteresis was inspected. Here we address the question how FMR \cite{FaSi13} spectra change based on the arrangement of such nanocubes. The spectra of resonant absorption are calculated which could be compared to single particle FMR spectra which may be recorded in new so-called micro resonator set-ups \cite{BaNa11,FaSi13}. We analyze the results in view of FMR performed in a microresonator \cite{NaSt05} (probing area around 20 $\mu$m) \cite{Terw12}. The latter allows for probing of significantly lower number of spins ($10^6$), whereas the conventional cavity setup detects the signal of minimum $10^{11}$ spins. The FMR measured in the microresonator gives thus a better link between the experimental and our simulated results.

\begin{figure}[!t]
\centering
\includegraphics[width=0.6\textwidth]{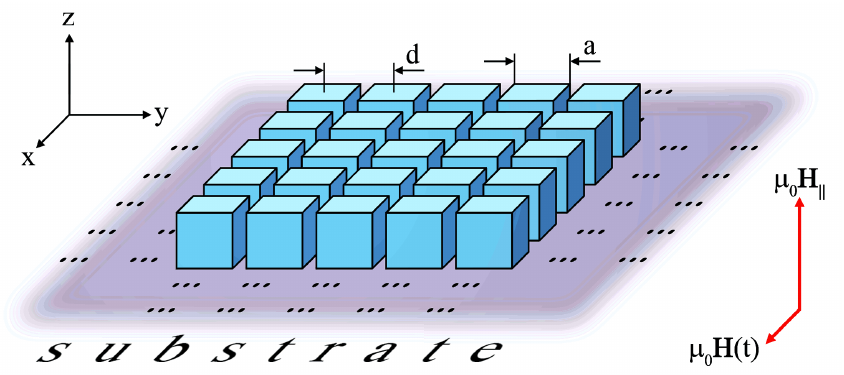}
\caption{Schematic view of the nanoparticles (example of a $5\times 5$-array) and mutual alignment of magnetic fields used in the simulations.}
\label{fig_1}
\end{figure}

\section{Theoretical model}

In the first approximation the FM nanocubes can be modeled as single domain classical nanoparticles with a huge magnetic moment (\textit{macrospin} \cite{StWo48,Brow78}). Hereby, the free energy for a single nanocube should include
\begin{equation}
\displaystyle F=F_{\mathrm{ANI}}+F_{\mathrm{ZMN}},
\label{eq_0}
\end{equation}
whereas the first term $F_{\mathrm{ANI}}=F_{\mathrm{MCA}}+F_{\mathrm{SHP}}$ contains contributions from the magnetocrystalline anisotropy (cubic for bcc-iron) rotated around the z-axis by angle $\phi_0$ \cite{remark_ca}
\begin{eqnarray}
\displaystyle F\ind{MCA}(\phi_0)=& \left(\frac{K_4}{M_{\mathrm{S}}^4}+\frac{K_6}{M_{\mathrm{S}}^6}M_{\mathrm{z}}^2\right) \left(\frac{1}{2}\left(M_{\mathrm{x}}^2-M_{\mathrm{y}}^2\right) \sin2\phi_0+M_{\mathrm{x}}M_{\mathrm{y}}\cos2\phi_0\right)^2 \nonumber \\
&+\frac{K_4}{M_{\mathrm{S}}^4}M_{\mathrm{z}}^2\left(M_{\mathrm{x}}^2+M_{\mathrm{y}}^2\right)
\label{eq_1}
\end{eqnarray}
as well as the shape anisotropy
\begin{equation}
F_{\mathrm{SHP}}=\frac{1}{4}\mu_0 M^2_{\mathrm{S}}(1-3N)\Big|_{N=\frac{1}{3}}=0,
\label{eq_2}
\end{equation}
in approximation of spheres \cite{foot_1} for the nanocubes. The Zeeman-contribution is given by
\begin{equation}
\displaystyle F_{\mathrm{ZMN}}=-\mu_0 \vec{M}\cdot \vec{H}\ind{\Sigma}(t),
\label{eq_3}
\end{equation}
where $\vec{H}\ind{\Sigma}(t)=\vec{H}_{||}+\vec{H}(t)$ is the total magnetic field.

For an ensemble of interacting nanoparticles the dipole-dipole interactions (DDIs) \cite{Jack75} between all the nanoparticles within the microresonator
\begin{equation}
\displaystyle F\ind{DDI}=\frac{\mu_0}{4\pi}\sum_{i\neq j}\left[\frac{\vec{M}_i\cdot \vec{M}_j-3(\vec{M}_i\cdot \vec{e}_{ij})(\vec{e}_{ji}\cdot \vec{M}_j)}{n^3_{ij}}\right]
\label{eq_4}
\end{equation}
are unavoidable, since the nanocubes are typically less than $100$~nm apart ($\mu_0=4\pi \cdot 10^{-7}$~[Vs/(Am)], $n_{ij}=r_{ij}/d_{\mathrm{min}}$, where $d_{\mathrm{min}}=40$~[nm] is the smallest possible inter-particle distance). Therefore, in the case of a nanoparticles' array eq. (\ref{eq_0}) should be augmented by eq. (\ref{eq_4}).

Numerically, the absorbed power $P\ind{FMR}(B)$ at $T=0$~[K] is calculated according to \cite{Usad06,SuUs08}
\begin{equation}
\displaystyle P\ind{FMR} \sim \sum_i \frac{1}{N\ind{T}T} \int_0^{N\ind{T}T}\mu_0 \vec{M}_i(t)\cdot \frac{\partial \vec{H}\ind{\Sigma}(t)}{\partial t}dt
\label{eq_5}
\end{equation}
which is proportional to the imaginary part of the transverse magnetic susceptibility $\chi\ind{im}$ \cite{Vons66}\textcolor{black}{ , where $N_{\mathrm{T}}$ denotes the number of averagings ($N_{\mathrm{T}}=1$ in the present calculations) for the period $T=2\pi/\omega$ of the external field $H(t)$.}

All parameters are taken as for bulk iron. Nanoparticles are modeled as macrospins with the saturation magnetization $M_{\mathrm{S}}=1.7\cdot 10^6$~[A/m] \cite{Coey10}. The size of such nanocubes is taken as $a^3=40\times 40\times 40$~[nm$^3$] \cite{Terw12}. The core-shell structure and its related varying material parameters are neglected for simplicity.

Anisotropy constants corresponding to the four-fold symmetry are $K_{\mathrm{4}}=4.8\cdot 10^4$~[J/m$^3$] and $K_{\mathrm{6}}=-1.0\cdot 10^4$~[J/m$^3$] \cite{Coey10}. The main anisotropy axes coincide with the coordinate system of the sample, i.e. the z-direction  (one of the easy axes) is the direction of applied static field ($\vec{B}=\mu_0\vec{H}_{||}=B\vec{e}\ind{z}$). The rf-field (oscillating magnetic field $\mu_0\vec{H}(t)=\vec{B}(t)=B_0\cos \omega t \vec{e}\ind{x}$ with the amplitude $B_0/\mu_0=2.2\cdot 10^3$~[A/m]) is applied in the plane of the nanocubes along the x-axis (cf. Fig. \ref{fig_1}).

The distance between the nanoparticles $r_{ij}$ (center-to-center) is inspected as the main parameter influencing the calculations. By default it is chosen as $d=60$~[nm] (Figs. \ref{fig_3}, \ref{fig_4}, \ref{fig_5}). According to the experimental results \cite{Terw12}, the frequency of the rf-field is set to $\omega/(2\pi)=9.5\cdot 10^9$~[Hz]. The ferromagnetic damping parameter which is crucial for the magnetization dynamics as well as for the FMR-spectra is chosen as $\alpha_{\mathrm{FM}}=0.02$ \cite{DoFi97}, \textcolor{black}{ which is larger than the one of bulk Fe, but very close to the value (0.03) used for successfully fitting a two-dimensional arrangement of a powder of Fe/Fe$_{\mathrm{x}}$O$_{\mathrm{y}}$ nanoparticles (Ref. \cite{TrMe08})}.

The magnetization of each FM nanocube $\vec{M}_i$ at time $t$ is found from the propagation of the Landau-Lifshitz-Gilbert \textcolor{black}{ (LLG)} \cite{LaLi35,Gilb55} equation of motion
\begin{eqnarray}
\displaystyle \frac{d\vec{M}_i}{dt} = && - \frac{\gamma}{1+\alpha^2_{\mathrm{FM}}} \left[\vec{M}_i\times \vec{B}_i^{\mathrm{eff}}(t)\right] \nonumber \\
&&- \frac{\alpha\ind{FM} \gamma}{1+\alpha^2_{\mathrm{FM}}}\frac{1}{M\ind{S}} \left[\vec{M}_i\times\left[\vec{M}_i\times \vec{B}_i^{\mathrm{eff}}(t) \right]\right],
\label{eq_5a}
\end{eqnarray}
where $\alpha\ind{FM}$ is the Gilbert damping and the local effective field is defined as $\vec{B}_i^{\mathrm{eff}}(t)=-\frac{\delta F}{\delta \vec{M}_i}$ at zero Kelvin. \textcolor{black}{ We also note that in the present calculations we disregard the effect of finite temperatures. The justification of this assumption follows from the comparison of the bulk parameters like saturation magnetization or the values of the anisotropy constants at zero Kelvin and at room temperature \cite{Ohan00}. In addition, in view of the blocking temperature of about $230$~K for the core/shell Fe/Fe$_{\mathrm{x}}$O$_{\mathrm{y}}$-nanocubes with averaged sizes of $18$ nm \cite{ShRo07}, we expect the superparamagnetic behaviour to be less significant for the sizes considered here.}

To achieve the required accuracy of the magnetization dynamics we used the time step of $0.1$ ps for solution of the LLG equation. As the damping parameter $\alpha_{\mathrm{FM}}$ for iron is quite small, it was necessary to restore the temporal dynamics of the system for numerous cycles of variable field to reach the stable precession mode. With this in mind, we skipped 99 cycles of $H(t)$ and used the magnetization dynamics recorded for the $100$th-cycle to calculate the FMR spectrum.

\textcolor{black}{ The LLG equation is solved numerically using the Heun method which is usually employed for solving the stochastic version of the LLG and converges in quadratic mean to the solution when interpreted in the sense of Stratonovich \cite{Nowa01}. }

The necessity to employ complex numerical calculations is dictated by intricate and nonlinear contributions of the DDIs following from eq. (\ref{eq_4}).

\section{Results}

\begin{figure}[h]
\centering
\includegraphics[width=0.45\textwidth]{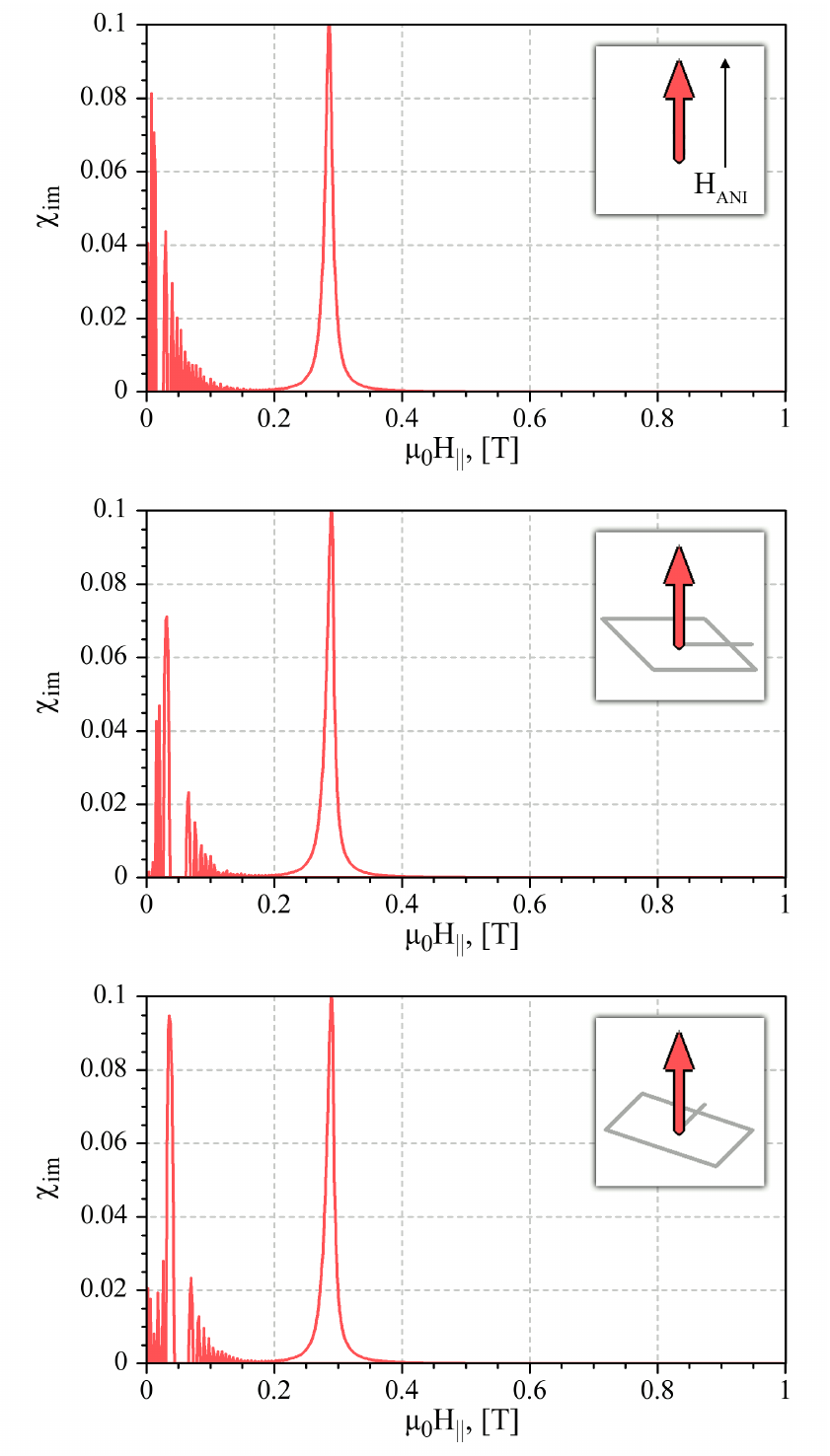}
\caption{Dependence of FMR spectra of a single nanoparticle (40x40x40 nm$^3$) on the crystalline anisotropy for an uniaxial anisotropy ($K_2=4.8\cdot 10^4$~[J/m$^3$]) directed along the z-axis (upper panel), cubic anisotropy ($K_4=4.8\cdot 10^4$~[J/m$^3$], $K_6=-1.0\cdot 10^4$~[J/m$^3$]) \cite{Coey10} with the axis parallel to the x-axis (central panel) and cubic anisotropy rotated by $\phi_0=\pi$/4 around the z-axis. The position of the principal peak is 0.286 T for the uniaxial case and 0.290 T for the case of cubic anisotropy.}
\label{fig_2}
\end{figure}

\subsection{Single-nanoparticle case}

First, we consider the FMR response from a single-particle, which can be directly compared to analytical expressions (\ref{eq_5}) for $i=1$. We performed the calculations using the material parameters provided above and applying magnetic fields as discussed in Sec. II.

The particle with a uniaxial anisotropy displays a pronounced peak at the given field  frequency $\omega/(2\pi)=\nu$ assuming a saturated magnetization ($M\ind{z}\approx M\ind{S}$). The position of this peak can be estimated from
\begin{equation}
\displaystyle B\ind{res}(\mathrm{uniax. anis.})=\frac{\omega}{\gamma}-\frac{2K\ind{2}M\ind{S}}\approx 0.28 [\mathrm{T}],
\label{eq_6}
\end{equation}
which results from the minimization of expression (\ref{eq_0}) for $K\ind{6}=0$, $\phi_0=0$. As inferred from Fig. \ref{fig_2} (upper panel) the simulated resonance peak at $\mu_0H_{||}$ = 0.286 [T] coincides well with the value predicted by eq. (\ref{eq_6}). For consideration of iron particles one should take into account their cubic anisotropy characterized by non-zero constants $K_4$ and $K_6$ in eq. (\ref{eq_1}). As one can see from the figure (cf. central and lower panels of Fig. \ref{fig_2}), the position of the main peak does not change significantly, moving for about 0.004 T towards higher fields. Rotation of the anisotropy landscape around the vertical axis by $\pi/4$ does not change the position of the main peak in full accordance with the analytical predictions.

For small values of static field ($\mu_0H<0.2$ [T]) the system can converge to different (meta-stable) precession modes, yielding a set of very narrow peaks (Fig. \ref{fig_2}). When the static field becomes large enough to reorient the magnetization of the particle along the z-axis, one can see only a single large peak associated with resonant absorption. To decrease the magnitude of the low-field peaks one can use two approaches. First, it is possible to discard a larger number of field cycles allowing the system to converge to the states with highest degree of stability. Indeed, in FMR-experiments the measurements are performed within microseconds or even seconds when steady trajectories are definitely reached. Therefore, in order to approach such prolonged waiting periods considerably surpassing nanosecond dynamics, we discard the initial 99 oscillation periods of the variable field. The other solution is to consider an array of nanoparticles, which, due to collective magnetization dynamics will effectively suppress the oscillation modes of a single particle, allowing to observe only magnetization configurations for which the entire system will be ordered to a considerable degree.

\begin{figure*}[!t]
\centering
\includegraphics[width=0.9\textwidth]{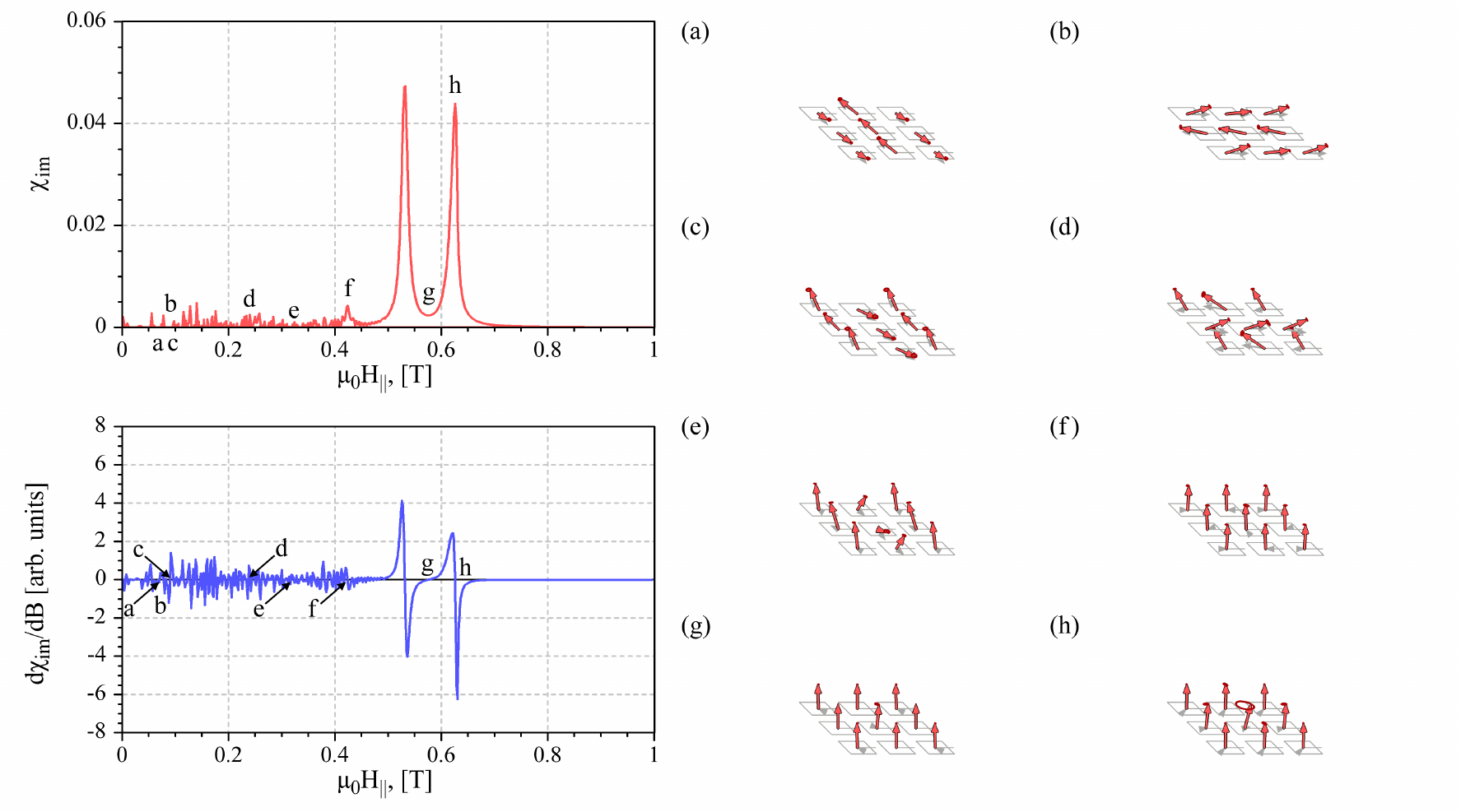}
\caption{FMR spectra of magnetic susceptibility and its first derivative with respect to the applied static field for the system of $3 \times 3$ cubic nanomagnets with the side 40~nm and distance between the centers of the particles $d$ = 60~[nm]. The strength of the DDI for the given $d$ is approximately $0.2$~T. The characteristic magnetic configurations are illustrated for the static field values $\mu_0H_{||}$: a) 0.056 T; b) 0.096 T; c) 0.1 T; d) 0.236 T; e) 0.322 T; f) 0.424 T; g) 0.58 T and h) 0.626 T. Arrows indicate the directions of the nanoparticles' macrospins. \textcolor{black}{ Closed curves marked at the tip of each macrospin indicate the steady-state precession shown for the last field cycle.}}
\label{fig_3}
\end{figure*}

\subsection{Nanoparticles' arrays}

\begin{figure*}[!t]
\centering
\includegraphics[width=0.9\textwidth]{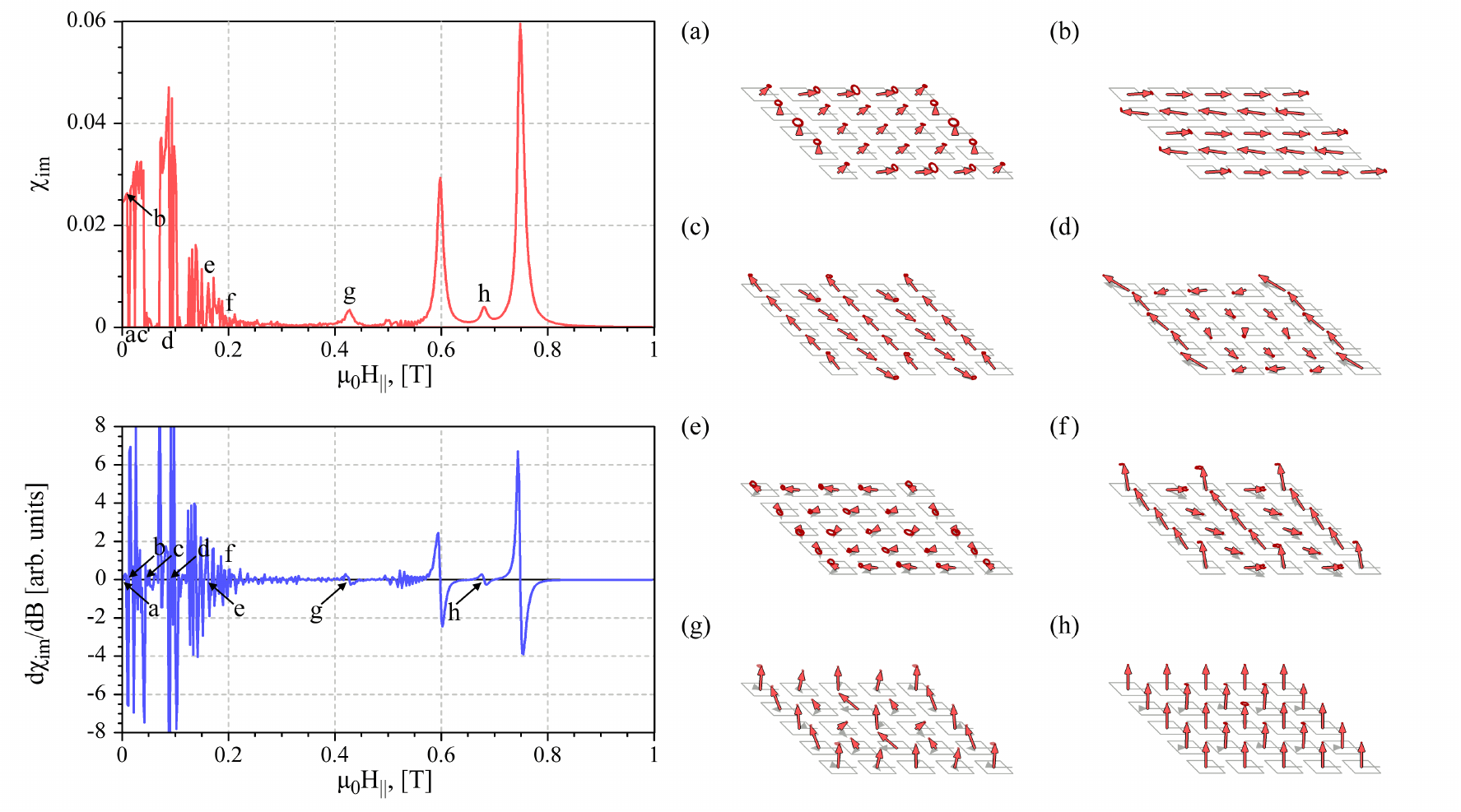}
\caption{FMR spectra of magnetic susceptibility and its first derivative for the system of $5 \times 5$ cubic nanomagnets with the side 40~nm and distance between the centers of the particles $d = 60$ [nm]. The characteristic magnetic configurations are illustrated for the static field values $\mu_0H_{||}$: a) 0 T; b) 0.012 T; c) 0.046 T; d) 0.092 T; e) 0.162 T; f) 0.2 T; g) 0.426 T and h) 0.68 T. Arrows indicate the directions of the nanoparticles' macrospins. \textcolor{black}{ Closed curves marked at the tip of each macrospin indicate the steady-state precession shown for the last field cycle.}}
\label{fig_4}
\end{figure*}

\begin{figure*}[!t]
\centering
\includegraphics[width=0.9\textwidth]{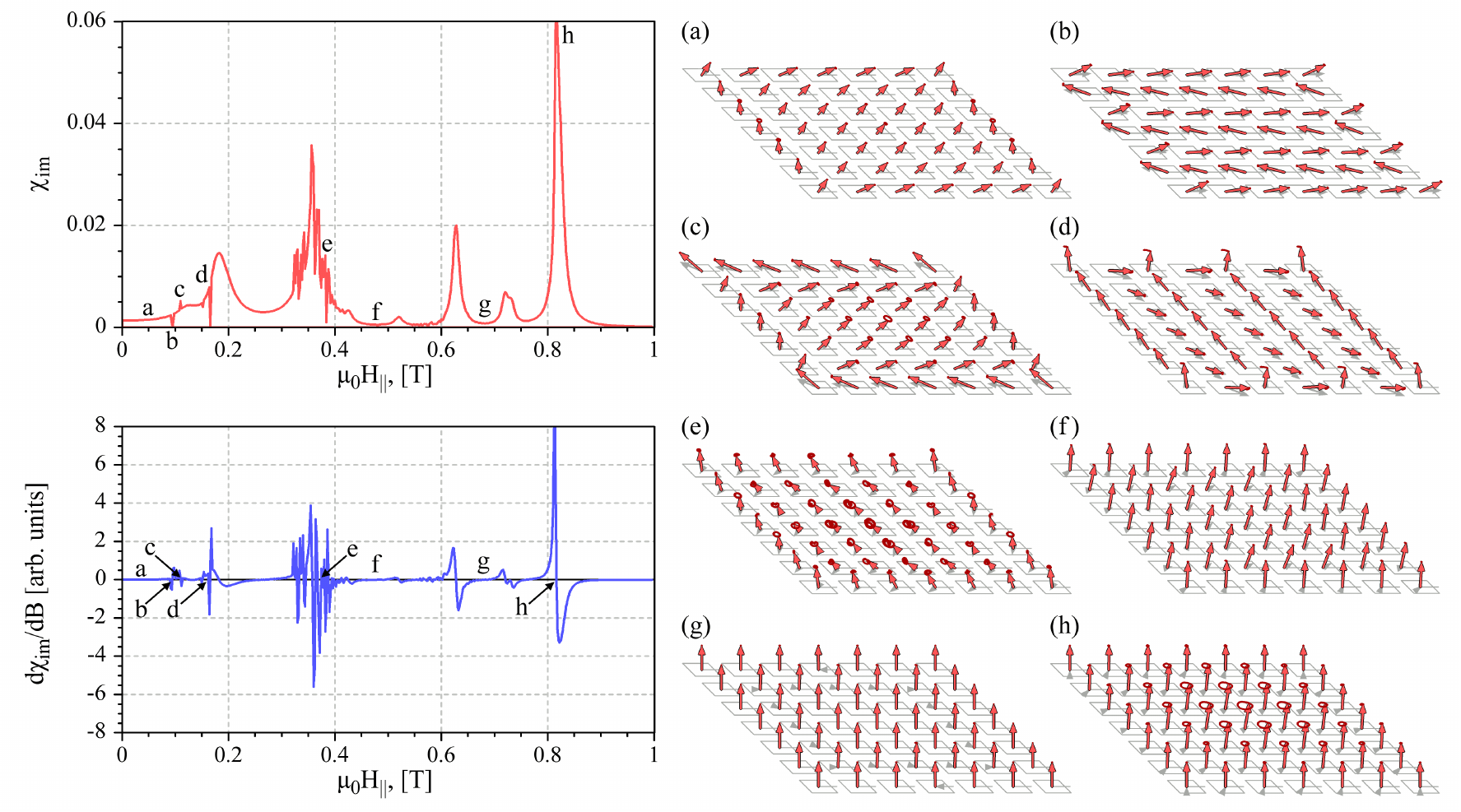}
\caption{FMR spectra of magnetic susceptibility and its first derivative for the system of $7 \times 7$ cubic nanomagnets with the side 40~nm and distance between the centers of the particles $d = 60$ [nm]. The characteristic magnetic configurations are illustrated for the static field values $\mu_0H_{||}$: a) 0.005 T; b) 0.096 T; c) 0.11 T; d) 0.166 T; e) 0.38 T; f) 0.468 T; g) 0.668 T and h) 0.818 T. Arrows indicate the directions of the nanoparticles' macrospins. \textcolor{black}{ Closed curves marked at the tip of each macrospin indicate the steady-state precession shown for the last field cycle.}}
\label{fig_5}
\end{figure*}

Further, we consider a system of non-overlapping cubic nanoparticles that can interact with each other via DDIs. The intensity of such interactions can be tailored by modifying the distance between the particles. It is natural to expect that the type of the collective modes taking place in the system will depend on the number of particles involved; therefore, we consider a complex optimization task with the aim to study the dependence of FMR-spectra on the number of particles and the distance between them in two-dimensional particle arrays. To illustrate the situation, we study the arrays formed by $3\times 3$ (Fig. \ref{fig_3}), $5\times 5$ (Fig. \ref{fig_4}) and $7\times 7$ (Fig. \ref{fig_5}) particles. All particles are considered perfectly aligned with $\phi_0 = 0$ and the distance between the centers of the particle is $60$~nm.

Even for the smallest $3 \times 3$ array the FMR spectrum becomes considerably different from a single-particle case (Fig. \ref{fig_3}). The most pronounced effect is the presence of two main peaks, which according to our calculations is caused by the dipole-dipole interaction between the particles. With variation of the static field, the system switches between several stable configurations. In the case when $\mu_0H_{||}$ is small, magnetization vectors of individual particles rather prefer the in-plane orientation (Fig. \ref{fig_3}, a-d). For stronger static fields the predominant particle's magnetization is oriented along the field. It is necessary to emphasize that different magnetic configurations require a certain energy for their formation, which may be associated with peaks or zeros of the FMR spectrum. However, it is not possible to identify the type of magnetic configuration from the FMR spectrum alone! Nevertheless, it is fruitful to follow the development of the system between different magnetic configurations that eventually lead to predominant alignment of magnetic moments for the case of high static field values. To simplify the discussion, we also plot in the figure the first derivative of the simulated FMR signal with respect to the static field, which will allow the easier detection of peak positions.

As one can see from the figure, at the field of 0.1 T magnetic moments of the particles are oriented along the cube's axis in an antiparallel fashion (Fig. \ref{fig_3}, a). The nucleation of such magnetic configuration occurs when the variable magnetic field pushes magnetization of the particles from their stable configuration and at the same moment the static field is weak enough to pull the magnetization out of the sample's plane. As x- and y-directions are equal for the cubic anisotropy, it is also possible to create the very similar arrangement along the other axis (Fig. \ref{fig_3}, b). With increasing static field the magnetization vectors of the particles continue to form antiparallel linear arrangements (Fig. \ref{fig_3}, c-e) with increasing $M_{\mathrm{z}}$ component signaling the increasing alignment with the static field. The dipole-dipole interaction has a stabilizing influence on the system; this can be seen by the fact that the deviation from the in-plane configuration starts with magnetic moments of the particles located at the corners of the square (Fig. \ref{fig_3}, c), which have neighbors only at one side so that the torque generated by them is highly unbalanced and anisotropic. The particle in the center of the array keeps the in-plane orientation up to considerably higher fields (Fig. \ref{fig_3}, e). When the static field becomes dominating, the magnetic moments of the particles start to oscillate closely around $H_{||}$ (Fig. \ref{fig_3}, f-h). It is worth mentioning that larger absorption occurs when magnetization moments of the central particle and perimeter particles are not perfectly synchronized (Fig. \ref{fig_3}, f and h); in the opposite case, one obtains a minimum on the FMR curve (Fig. \ref{fig_3}, g) which is easily detectable on the derivative plot as the point where $d\chi_{im}/dB$ vanishes.

For the systems formed with a larger number of particles it is expectable to achieve a wider variety of possible states, which is definitely confirmed by our simulations (Figs. \ref{fig_4}, \ref{fig_5}). For the case of the square of $5 \times 5$ particles' array at zero applied field, the system is characterized by a "leaf" state (Fig. \ref{fig_4}, a). \textcolor{black}{ This effect has also been observed in the recent FMR-study \cite{RaBa13}}. Under small static field the system switches to anti-parallel arrangement of magnetic moments that can be directed along either axes in the plane of the cube's base (Fig. \ref{fig_4}, b, c). For higher fields the magnetic configuration starts to disorder from the outside (Fig. \ref{fig_4}, d), which can be explained by lower stability of the particles located at the perimeter of the array. It is important to emphasize that the modified "leaf" state can be achieved in the system under non-zero static field (Fig. \ref{fig_4}, e) with an obvious z-component of the magnetization signaling the magnetization alignment in the direction of the static field. Remarkably, not only corners of the FM particles' array may be unstable. For example, for the case of static field of 0.2 T (Fig. \ref{fig_4}, f) one can see the anti-parallel arrangement of magnetic moments with almost vertical orientation of the of magnetization vector for the particles located at the corners of the array and also in the middle of the perimeter of the array along the x-axis. This type of magnetic ordering is reminiscent of the formation of flux closure domains in solid ferromagnets, where the stray field magnetostatic energy is responsible for the formation of antiparallel orientations of magnetization vectors to lower the total energy of the system. It is therefore important to notice that a similar arrangement of magnetization vectors may also occur in an ordered array of discrete particles. The change from the predominant in-plane state to that parallel to the applied field can be estimated from the effective field that includes dipole-dipole interactions $H^{\mathrm{eff}}_{\mathrm{DDI z}}=-\frac{\delta F\ind{DDI}}{\delta M_{i \mathrm{z}}}\sim \frac{\mu_0 M\ind{S}}{4\pi}\approx 0.2$~[T]. Thus, when the static field exceeds the threshold of approximately $0.2$~T, it becomes possible to achieve the states when the magnetization reorients out of the sample plane. A complex non-collinear magnetization precession can be observed in the system for transitional fields (Fig. \ref{fig_4}, g). When static fields become high enough to be responsible for the formation of the main FMR peaks one notices that for a larger array size the first peak located at 0.6 T has about a half-height of the second one, located at about 0.75 T. This discrepancy of peak height will be manifested even more for larger arrays (Fig. \ref{fig_5}). To our opinion, this phenomenon can be explained by the fact that the peak located at the higher static fields corresponds to more pronounced ordering in a magnetic system that naturally needs more energy to be formed. Nevertheless, the precise definition of the effects responsible for the variation in height of the main FMR peaks requires a more detailed analysis that goes beyond the scope of the current paper and will be considered elsewhere. Comparing the FMR spectrum of a $5 \times 5$ array with the one of a $3 \times 3$ particles one immediately finds a secondary peak between the main peaks (Fig. \ref{fig_4}, h) that is characterized by a considerable ordering of magnetic moments in the system.

Increasing the array to $7 \times 7$ nano-particles, one can directly see that the magnetization dynamics of the system has reached a considerable stabilization so that multiple narrow peaks observable at low values of static field merge into far smoother spectral bands (Fig. \ref{fig_5}). The lowest stationary state of the system is still the "leaf" state (Fig. \ref{fig_5}, a), which can also appear in a "modified" configuration (Fig. \ref{fig_5}, c) when  the magnetic moments of the particles located at the perimeter of the array achieve a considerable component along the z-axis. The antiparallel arrangement of magnetic moments can be also observed for the large nano-particle array, with predominant magnetization orientation in the plane of the sample (Fig. \ref{fig_5}, b). Similarly to the situation illustrated for the case of the $5 \times 5$ particles array, upon increase of the static magnetic field the moments of the particles that has lower number of nearest neighbors may align almost parallel to the direction of $H_{||}$. An inhomogeneously broadened peak located at about 0.35 T is formed by narrow peaks corresponding to an asynchronous magnetization rotation, when perimeter and the center of the particle array respond somewhat differently to the action of the variable field (Fig. \ref{fig_5}, e). The two main peaks in the FMR spectrum undergo further modifications - the first peak located at 0.63 T becomes significantly lower while the second peak at 0.825 T remains high (Fig. \ref{fig_5}, h). The small secondary peak appearing in the middle of two aforementioned main peaks splits in two. The value of $\chi_{im}$ on both sides of this secondary peak group is rather low; the magnetic configuration of the system at the field value of 0.668 T is characterized with the high magnetization alignment along the z-axis.

\begin{figure*}[!t]
\centering
\includegraphics[width=0.9\textwidth]{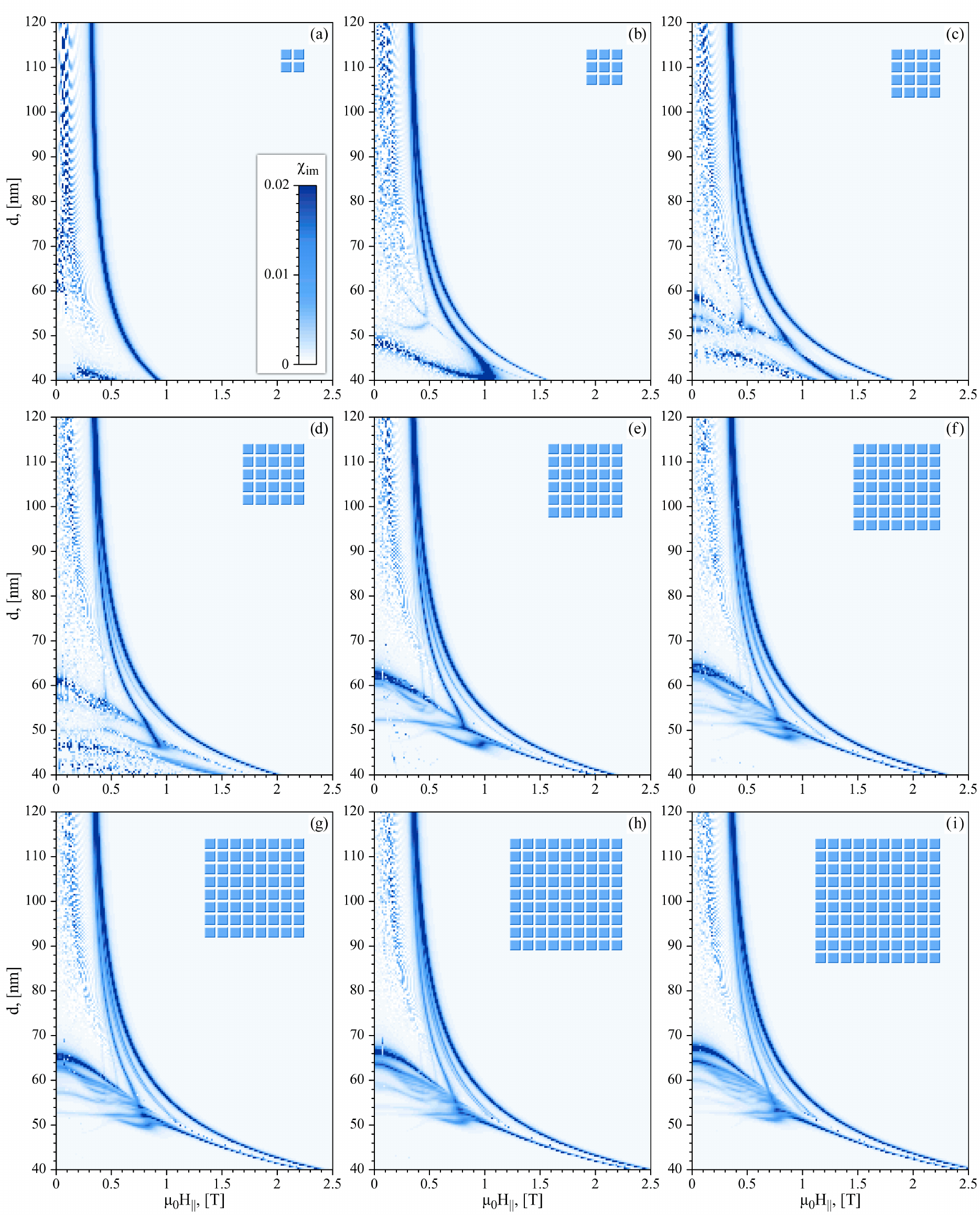}
\caption{Dependence of FMR spectra on inter-particle distance in the system of cubic magnetic particles 40x40x40 nm$^3$ in size for the arrays composed of: a) 2x2 particles; b) 3x3 particles, c) 4x4 particles; d) 5x5 particles; e) 6x6 particles; f) 7x7 particles; g) 8x8 particles; h) 9x9 particles and i) 10x10 particles. The schematic depiction of the corresponding particles' array is given at the upper right corner of each panel.}
\label{fig_6}
\end{figure*}

Therefore, numerical simulations of FM particle arrays composed with cubic nano-magnets with the side of 40 nm located in the nodes of the square grid with node-to-node distance of 60 nm show a rich variety of stable magnetic configurations at low static field values. Two main FMR peaks shift to the higher field values for the increasing size of the nanoparticles' array that can be explained by the need to apply stronger field to achieve magnetic alignment of the larger array. At the same time, it is not exactly clear how the magnetic response of the system will change with variation of the distance between particle centers $d$. To study this, it was decided to calculate two-dimensional plots displaying changes to FMR spectra with $d$ for different number of the particles forming the array, from $2 \times 2$ to $10 \times 10$ particle configurations. The results are shown in Fig. \ref{fig_6}. The values of $\chi_{im}$ are coded with color intensity, so that FMR peaks will correspond to dark bands while light color will correspond to a vanishing FMR signal. The color scale is given as an inset for Fig. \ref{fig_6}, a.

The main feature to all the figures is that with the increase of the distance between the centers of the particles to $d=$ 120 [nm] the main peaks of the FMR spectrum merge into a single peak located at approximately 0.4 T. The presence of a single peak is due to the fact that the dipole-dipole interaction between the particles becomes negligible, so that the collective magnetic motion in the system disappears and the individual particles become virtually non-interacting. Following the slope defining the peak position as a function of applied static field and the distance between the particles it is possible to estimate that for the values of $d > 160$ [nm] (that is, four times the dimension of the particle) the position of the main absorption peak will correspond to the value obtained for a single particle system (Fig. \ref{fig_2}). The richest set of FMR peaks and magnetic configuration takes place in the system with separation between the center of the particles up to 80 nm, which is twice the particle size. For larger particle arrays the available variety of magnetization states are wider, so that it is natural to expect that the fine structure of the FMR spectra will appear for the larger $H_{||}$. For simple systems such as $2 \times 2$ particles (Fig. \ref{fig_6}, a) no fine structure is essentially observed, so that the decrease of the inter-particle distance is associated only with the displacement of a single resonance peak to the right, reaching the value of $\mu_0H_{||} = 0.9$ [T] for the case when the particles are standing side-by-side ($d=$ 40 [nm]). Only in the case of the $3 \times 3$ particle array (Fig. \ref{fig_6}, b) the main resonance peak splits in two, giving rise to two states with magnetization aligned along the z-axis characterized with the different stability. The existence of two peaks indicates different absorption for the inner nano-particle and the outer ones.

\begin{figure*}[!t]
\centering
\includegraphics[width=0.9\textwidth]{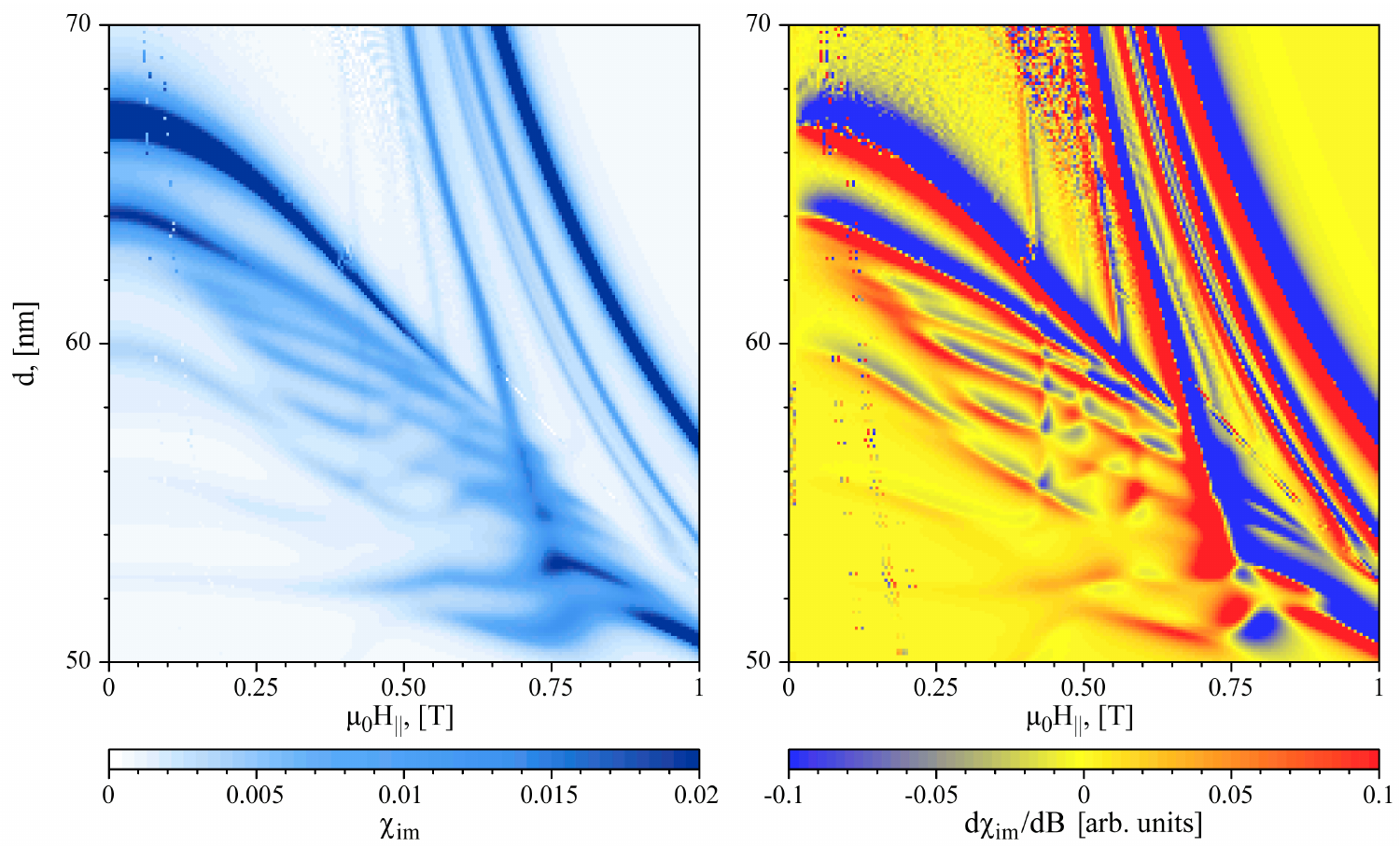}
\caption{Fine structure of FMR peaks for the case of 10x10 particle array for the variation of the inter-particle distance in the range $(50:70)$~nm . Left panel represents the susceptibility values and right panel - the derivative thereof.}
\label{fig_7}
\end{figure*}

A faint secondary peak can be detected between two main FMR maxima starting from the systems formed with $5 \times 5$ particles (Fig. \ref{fig_6}, d). For larger systems, this peak becomes wider and eventually splits (Fig. \ref{fig_6}, i). For the small values of static magnetic field $\mu_0H_{||} \lessapprox 0.3$ [T] the particle system becomes unstable that is witnessed by the visible noise on $\chi_{im}(\mu_0H_{||}, d)$ plots. This noise appears due to the fact that application of the variable magnetic field triggers the magnetization reordering in the system, which may take one of possible stable magnetic configurations -- such as the "leaf" state or antiparallel arrangements described before.

As the system can converge to the states with the preferred magnetization orientation along the x- and y-axes, it may require slightly different energies to produce the corresponding magnetization order. For that, the FMR spectrum will display peaks of different heights. At the same time, all these low-field peaks will be very narrow, reflecting the fact that the given magnetic configuration was essentially $triggered$ by the variable field. That is, running the simulations several times starting from random initial orientation of magnetic moments of the particles constituting the system, it will be possible to obtain $different$ magnetization configurations for the same magnitudes of static magnetic field and parameter $d$. Namely due to this intermittent property one can see a clear noise for the low-field area of the plots. However, it will be natural to expect that more pronounced stability of the system can be achieved in the case when the array will be made of larger number of the particles. Our simulations perfectly illustrate this suggestion, showing that starting with array sizes of $6 \times 6$ particles the low-field noise disappears for the distances between the centers of the particles below 60 nm (Fig. \ref{fig_6}, e). Further increase of the system size extends the noiseless area almost up to $d=75$ [nm] (Fig. \ref{fig_6}, i). These results, to our opinion, are extremely important for the development of device prototypes based on the arrays of nanoparticles, suggesting that large particle arrays (such as those composed by $10 \times 10$ elements) will feature considerably more stable magnetic configurations that will have a good potential to withstand thermal fluctuations. Also, larger particle arrays display a much richer set of magnetic configurations at low static fields, that can be a useful property if one will find the way of controllable switching between the desired magnetic configurations.

To better illustrate the fine structure of the FMR spectra for a complex system, we have calculated a detailled close-up of the FMR spectra for the system composed of $10 \times 10$ particles (Fig. \ref{fig_7}). The figure clearly shows the large intensity of two peaks - the peak located at $\mu_0H_{||} > 0.6$ [T] and the peak $\mu_0H_{||} < 0.5$ [T] attainable for the array grid parameter $d$ varying in the ranges of 62-68 nm. The set of smaller secondary peaks can be observed for closer inter-particle distances and higher static fields. It is important to emphasize that the results of our simulations can be used to fine-tune the parameters of a nano-particle array -- such as inter-particle distance -- in order to achieve a better magnetic stability of the system or access a wider spectrum of possible magnetic configurations. If it will be desirable to find the exact peak position one can consult the figure representing the derivative of the susceptibility function $d\chi_{im}(\mu_0H_{||},d)/dB$, presented in the right panel of Fig. \ref{fig_7}.

\section{Conclusions}

The present numerical calculations performed within the macrospin approximation for bulk parameters of iron nanocubes support the crucial importance of the long range ferromagnetic dipole-dipole interactions on the spectra of absorbed power probed by means of FMR. Our main finding is that depending on the number of the nanocubes regularly placed within the array, the FMR spectra may show multiple absorption peaks below a certain inter-particle distance in the range of $80\div 100$~nm (Fig. \ref{fig_6}). Above this threshold the maxima of the FMR absorption merge into a single peak imitating the behaviour of a single-nanoparticle spectrum. Among all energy contributions entering into eqs. (\ref{eq_0}) and (\ref{eq_4}) the most relevant are the DDI- and the Zeeman-contributions. The effective field induced by the anisotropy scales only as $\frac{2K_4}{M\ind{S}}\approx 0.05$~[T], whereas the same induced by the DDIs is below $\frac{\mu_0 M\ind{S}}{4\pi}\approx 0.2$~[T]. As a consequence of this predominance we observe an appearance of further resonance peaks and shifting of them towards higher magnetic fields (e.g. Fig. \ref{fig_5} above 0.5 T, vs. Fig. \ref{fig_2}) and witness numerous low fields absorptions related to different in-plane magnetization configurations (Fig. \ref{fig_5}, (a)-(d)).

We note that the existence of numerous low-field peaks observed in Figs. \ref{fig_2}, \ref{fig_3}, \ref{fig_4} and partly \ref{fig_6} is explained by a non-equilibrium situation created by a combination of weak static and rf-fields when calculated at zero Kelvin without averaging. \textcolor{black}{ The non-equilibrium state has its origin in the initial state of the magnetization to be along the y-axis, which is one of the ground states dictated by the DDIs. For the initial state along the z-axis, the low-field peaks for a single-particle case, e.g. as those shown in Fig. \ref{fig_2}, are significantly suppressed.} In the experimental situation performed e.g. at room temperature, such effects typically vanish due to a relatively high measurement time (several seconds) resulting in a vast averaging rate.

Elevated temperatures can significantly modify the obtained results. In particular, as was shown earlier \cite{SuUs08} for ensembles of single-domain FM nanoparticles all having randomly oriented easy anisotropy axes, the FMR-spectra strongly broadened at low temperatures become narrow and shift towards higher static fields at temperatures close to the room temperature. We expect a similar effect for the array of the nanocubes studied here.

\begin{figure*}[!t]
\centering
\includegraphics[width=0.7\textwidth]{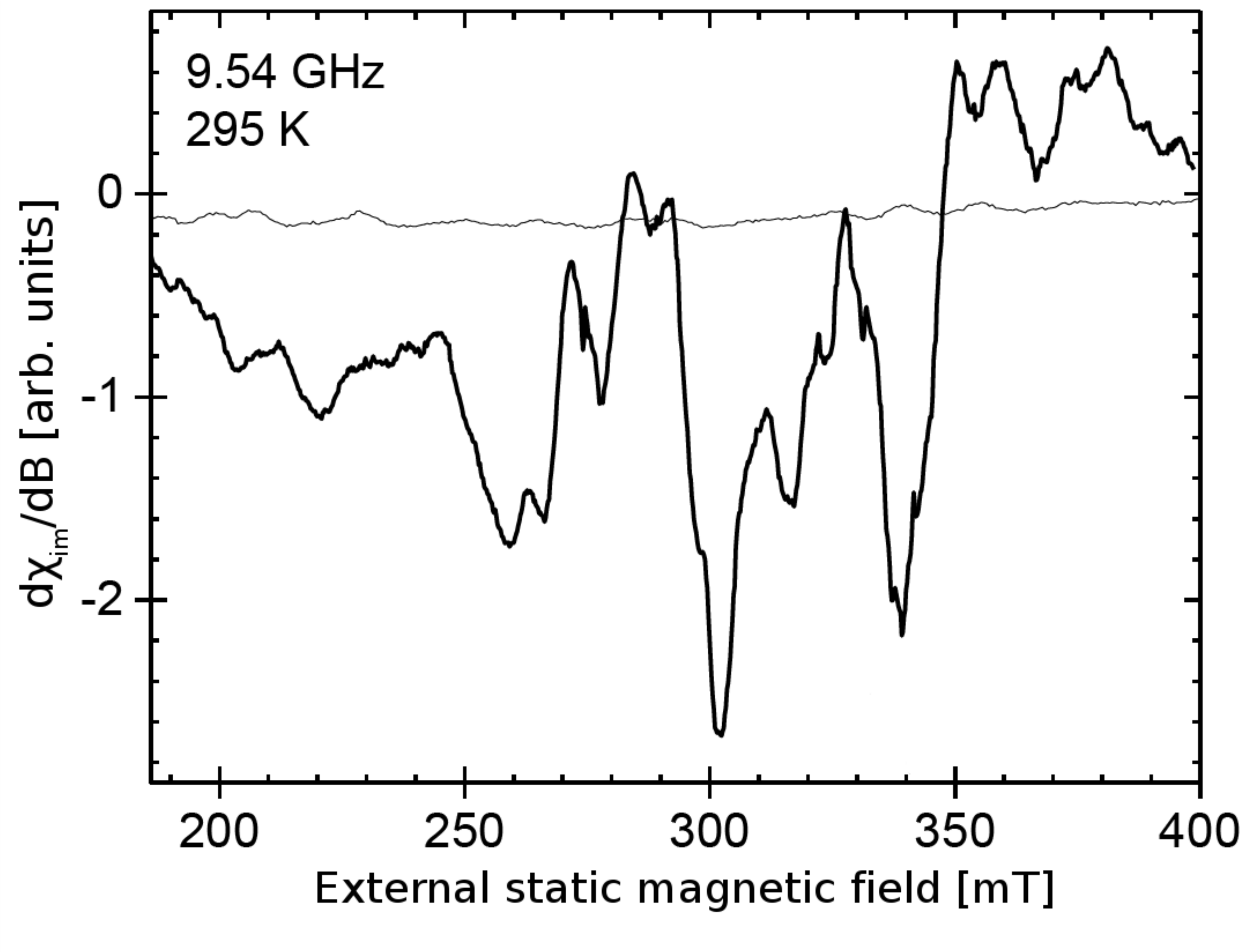}
\caption{\textcolor{black}{ Experimental FMR spectrum (thick line) of 24 Fe/Fe$_{\mathrm{x}}$O$_{\mathrm{y}}$-nanocubes recorded at 9.54 GHz at room temperature. The edge length of the cubes is 43 nm. The spectrum is the result of 24 overlapping resonances from individual cubes which have different resonance fields due to small changes in magnetic anisotropy energy, orientation and different dipolar coupling strengths between clusters of cubes. The thin line shows the background signal of the empty microresonator.}}
\label{fig_8}
\end{figure*}

\textcolor{black}{ Finally, our simulations may be compared with preliminary results (Fig. \ref{fig_8}) \cite{Terw12} for the resonant response of 24 Fe/Fe$_{\mathrm{x}}$O$_{\mathrm{y}}$-nanocubes randomly placed on a substrate. In particular, the FMR-spectrum of perfectly ordered 25 nanocubes (Fig. \ref{fig_4}) with center-to-center interparticle distance of 60 nm suggests two main absorption peaks located at $0.6$ and $0.75$~T. When the results of distance-dependent FMR are taken into account, e.g. from Fig. \ref{fig_6}, d), we observe a shift and a merging of them towards lower $B$-fields. This, and the fact that some of the isolated nanocubes can give a significant "own" contributions \cite{unpub}, can reproduce the experimentally observed results (Fig. \ref{fig_8}). A quantitative comparison, however, is beyond the scope of the present publication, since the parameter space for fitting the experimental spectrum contains too many unknown values for the shape anisotropy, oxidation state, separation and orientation of the 24 nanocubes.}

\section{Acknowledgements}

The authors gratefully acknowledge the funding by the grants of the DFG SU 690/1-1 and FA 209/15-1 (Germany) and of the CONACYT as Basic Science Project 129269 (Mexico).


\begin{thebibliography}{32}
\bibitem{remark_ca}
Expr. (\ref{eq_1}) can be derived from the standard relation for the cubic anisotropy energy density, i.e. $\frac{K_4}{M_{\mathrm{S}}^4}\left(M_{\mathrm{x}}^2M_{\mathrm{y}}^2+M_{\mathrm{y}}^2M_{\mathrm{z}}^2+M_{\mathrm{x}}^2M_{\mathrm{z}}^2\right)+\frac{K_6}{M_{\mathrm{S}}^6} M_{\mathrm{x}}^2M_{\mathrm{y}}^2M_{\mathrm{z}}^2$ \cite{Coey10}, when it is multiplied by the rotational matrix representing a rotation of the system (Fig. \ref{fig_1}) by the angle $\phi_0$. In other words, angle $\phi_0$ denotes a rotation of the cube's side relatively to the x-axis. Note, that we choose the nomenclature of Ref. \cite{Farl98} in which the leading cubic anisotropy term is defined as $K_4$ (instead of $K_1$) to avoid confusion with leading uniaxial anisotrpy term $K_2$ which has been defined as $K_1$ in the literature. The subscripts in Farle's notation reflect the symmetry of the anisotropy term.
\bibitem{foot_1}
It should be noted, that FM-nanoparticles having cubic form possess in general a non-zero demagnetizing field, which grows with incresing nanoparticle's size\cite{BeGr03,BeGr05}.
\end{thebibliography}
\end{document}